\documentclass[twocolumn,superscriptaddress,showpacs,preprintnumbers,amsmath,amssymb,doublespacing,prl]{revtex4-1}

\usepackage[dvipdfmx]{graphicx}
\usepackage{bm}

\newcommand{\msr}{$\mu$SR}
\newcommand{\cdo}{Cd$_2$Os$_2$O$_7$}
\newcommand{\cdoro}{Cd$_2$(Os$_{1-x}$Re$_{x}$)$_2$O$_7$}

\newcommand{\sro}{Sr$_2$IrO$_4$}

\begin{document} 

\title{Coupled Spin-Charge-Phonon Fluctuation in the All-In/All-Out  Antiferromagnet \cdo} 

\author{A. Koda}
\affiliation{Muon Science Laboratory and Condensed Matter Research Center, Institute of Materials Structure Science, High Energy Accelerator Research Organization (KEK), Tsukuba, Ibaraki 305-0801, Japan}
\affiliation{Department of Materials Structure Science, The Graduate University for Advanced Studies, Japan}
\author{H. T. Hirose}\thanks{Present address: National Institute for Materials Science, Tsukuba, Ibaraki 305-0003, Japan}
\affiliation{Institute for Solid State Physics, University of Tokyo, Kashiwa, Chiba 277-8581, Japan}
\author{M. Miyazaki}\thanks{Present address: Muroran Institute of Technology, Muroran, Hokkaido 050-8585, Japan}
\affiliation{Muon Science Laboratory and Condensed Matter Research Center, Institute of Materials Structure Science, High Energy Accelerator Research Organization (KEK), Tsukuba, Ibaraki 305-0801, Japan}
\author{H. Okabe}
\affiliation{Muon Science Laboratory and Condensed Matter Research Center, Institute of Materials Structure Science, High Energy Accelerator Research Organization (KEK), Tsukuba, Ibaraki 305-0801, Japan}
\affiliation{Department of Materials Structure Science, The Graduate University for Advanced Studies, Japan}
\author{M. Hiraishi}
\affiliation{Muon Science Laboratory and Condensed Matter Research Center, Institute of Materials Structure Science, High Energy Accelerator Research Organization (KEK), Tsukuba, Ibaraki 305-0801, Japan}
\author{I. Yamauchi}\thanks{Present address: Department of Physics, Graduate School of Science and Engineering, Saga University, Saga 840-8502, Japan}
\affiliation{Muon Science Laboratory and Condensed Matter Research Center, Institute of Materials Structure Science, High Energy Accelerator Research Organization (KEK), Tsukuba, Ibaraki 305-0801, Japan}
\author{K. M. Kojima}\thanks{Present address: Centre for Molecular and Materials Science, TRIUMF, Vancouver, BC V6T2A3, Canada}
\affiliation{Muon Science Laboratory and Condensed Matter Research Center, Institute of Materials Structure Science, High Energy Accelerator Research Organization (KEK), Tsukuba, Ibaraki 305-0801, Japan}
\affiliation{Department of Materials Structure Science, The Graduate University for Advanced Studies, Japan}
\author{I. Nagashima}
\affiliation{Institute for Solid State Physics, University of Tokyo, Kashiwa, Chiba 277-8581, Japan}
\author{J. Yamaura}\thanks{Present address: Materials Research Center for Element Strategy, Tokyo Institute of Technology, Yokohama, Kanagawa 226-8503, Japan}
\affiliation{Institute for Solid State Physics, University of Tokyo, Kashiwa, Chiba 277-8581, Japan}
\author{Z. Hiroi}
\affiliation{Institute for Solid State Physics, University of Tokyo, Kashiwa, Chiba 277-8581, Japan}
\author{R. Kadono}\thanks{e-mail: ryosuke.kadono@kek.jp}
\affiliation{Muon Science Laboratory and Condensed Matter Research Center, Institute of Materials Structure Science, High Energy Accelerator Research Organization (KEK), Tsukuba, Ibaraki 305-0801, Japan}
\affiliation{Department of Materials Structure Science, The Graduate University for Advanced Studies, Japan}

%

\begin{abstract}
We report on a novel spin-charge fluctuation in the all-in-all-out pyrochlore magnet \cdo, where the spin fluctuation is driven by the conduction of thermally excited electrons/holes and associated fluctuation of Os valence. The fluctuation exhibits an activation energy significantly greater than the spin-charge excitation gap and a peculiar frequency range of $10^{6}$--$10^{10}$ s$^{-1}$.  These features are attributed to the hopping motion of carriers as small polarons in the insulating phase, where the polaron state is presumably induced by the magnetoelastic coupling via the strong spin-orbit interaction.  Such a coupled spin-charge-phonon fluctuation manifests as a part of the metal-insulator transition that is extended over a wide temperature range due to the modest electron correlation comparable with other interactions characteristic for 5$d$-subshell systems.
\end{abstract}

\maketitle 

\section{I. Introduction}
The metal-insulator (MI) transition has been serving as hunting ground for a wealth of emergent exotic properties in 3$d$ transition-metal compounds, where the role of electronic correlation ($U$) is the central topics of the Mott insulators. The current expansion of research to the 4$d$ and 5$d$ transition metal oxides is adding a new avenue to the relevant field, where the novel electronic phases induced by spin-orbit (SO) interaction comparable to $U$ are under much discussion for Ir and Os pyrochlore compounds \cite{Pesin:10,Kargarian:11,Imada:14,Schaffer:16}.  Recently, the effects of the strong SO interaction on the magnetic properties in these compounds are attracting growing attention.  The SO interaction mediates coupling between spin and lattice degrees of freedom, leading to a peculiar magnetic structure concomitant with the lattice modulation associated with the MI transition. The magnetism in the perovskite magnet \sro\ comprises a typical example, where the iridium spin structure is aligned with the rotation of IrO$_6$ octahedra via the Dzyaloshinskii-Moriya interaction \cite{Kim:08,Kim:09}.  

The very recent revelation of the strong spin-phonon coupling in \cdo\ put this compound in the spotlight as an excellent stage for investigating the unconventional effect of the SO interaction on magnetism \cite{Sohn:17}. Previously, \cdo\ was known as a unique example in that the MI transition of purely electronic origin (without change in the crystal symmetry) was suggested by a modest increase of electrical resistivity ($\rho$) at 225--6 K \cite{Sleight:74,Mandrus:01}. Magnetic susceptibility and specific heat measurements suggested the conventional N\'eel order below $T_N\simeq227$ K, which was in line with the Slater mechanism for the MI transition \cite{Mandrus:01,Slater:51}. However, the revelation of a non-collinear all-in-all-out (AIAO) spin structure (i.e., the propagation vector ${\bm q} = 0$) implied by resonant X-ray magnetic scattering (RXMS)  \cite{Yamaura:12,Shinaoka:12} and subsequent neutron diffraction measurements \cite{Calder:16} ruled out the Slater mechanism, leading to the suggestion that the MI transition was induced by the variation of Fermi surface topology (the Lifshitz transition) \cite{Yamaura:12,Hiroi:15}.

\begin{figure}[tb]
	\begin{center}
\includegraphics[width=0.95\linewidth]{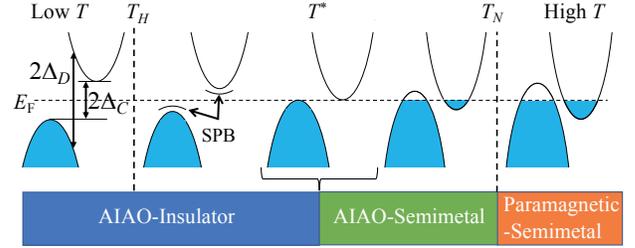}
			\caption{Schematic diagrams of the band structure and associated electronic/magnetic phases based on optical spectroscopy in \cdo\ \cite{Sohn:15}. Magnetic transition [with All-In/All-Out (AIAO) spin structure] occurs in the midst of semimetal phase at $T_N$, whereas the semimetal-to-insulator transition occurs in the AIAO-magnet phase at $T^*\simeq 210$ K.  In this work, the small-polaron band (SPB) is suggested to convey thermally activated carriers for $T_H\lesssim T\lesssim T^* $. $\Delta_C$ ($\Delta_D$) and $E_F$ refer to the indirect (direct) band gap in the insulating phase and the Fermi level, respectively.}
			\label{phasdia}
	\end{center}
\end{figure} 

Meanwhile, as shown in Fig.~\ref{phasdia}, it was inferred from optical spectroscopy experiment that a true charge gap ($\Delta_C$) opens at around $T^*\simeq210$ K ($< T_N$), indicating that the relevant phase is split into the AIAO-semimetal (AS) and AIAO-insulator (AI) phases \cite{Sohn:15}. While the observed change in the spectral weight is consistent with that predicted by the LDA+$U$ calculation, the discrepancy between $T_N$ and $T^*$  suggests a successive MI transition that may accompany a variety of unconventional critical behaviors. The issue seems also related with anomalous damping of muon spin rotation (\msr) signal between $T_H\simeq150$ K and $T_N$ in the AIAO-magnet phase reported in the earlier study \cite{Koda:07}.  

Here, we reveal the presence of novel spin-charge-phonon fluctuation lurking under the guise of quasistatic AIAO magnetic order. The spin fluctuation is driven by the conductive motion of electrons/holes that accompanies the valence fluctuation of Os ions throughout both the AS and AI phases over a temperature range between $T_H$ and $T_N$, which is further corroborated by the enhancement of spin fluctuation upon hole doping [by Re substitution for Os, \cdoro\ with $x=0.01$ and 0.03] below $T_H$ where the influence of intrinsic carriers is negligible.  The activation energy for the fluctuation is significantly greater than that for the spin-charge excitation, suggesting that the carrier mobility is limited by the magnetoelastic coupling via the SO interaction \cite{Sohn:17}.  We show that the scenario of small-polaron (called {\sl magnetoelastic} polaron) is quantitatively in line with the temperature dependence of the fluctuation rate, where $T_H$ represents the magnitude of the magnetoelastic coupling. 

\section{II. Experimental Details}

Single crystalline (sc-) samples of \cdo\ were prepared by the chemical vapor transport technique  \cite{Yamaura:12}, while polycrystalline samples of \cdoro\ were prepared by solid phase reaction method.  Conventional $\mu$SR measurements were performed on sc-\cdo\ at the Paul Scherrer Institute (PSI), Switzerland, and on the polycrystalline samples of \cdoro\ at TRIUMF, Canada, where a nearly 100\% spin-polarized beam of positive muons with a momentum of $\sim$27 MeV/c was irradiated to these samples loaded to a He-flow cryostat, and time evolution of muon spin polarization was monitored by recording positron decay asymmetry [$A(t)$] as a function of time. 
During the measurement under a zero field (ZF), residual magnetic field at the sample position was reduced below 10$^{-6}$~T with the initial muon spin direction parallel to the muon beam direction.   Measurements were also performed under a couple of different magnetic fields (0.2, 0.4 T) applied parallel to the initial muon polarization (longitudinal field, LF) to deduce the muon-Os hyperfine parameter ($\delta_\mu$) and its fluctuation rate ($\nu$) reliably. 

\section{III. Result}

\begin{figure}[t]
	\begin{center}
\includegraphics[width=\linewidth]{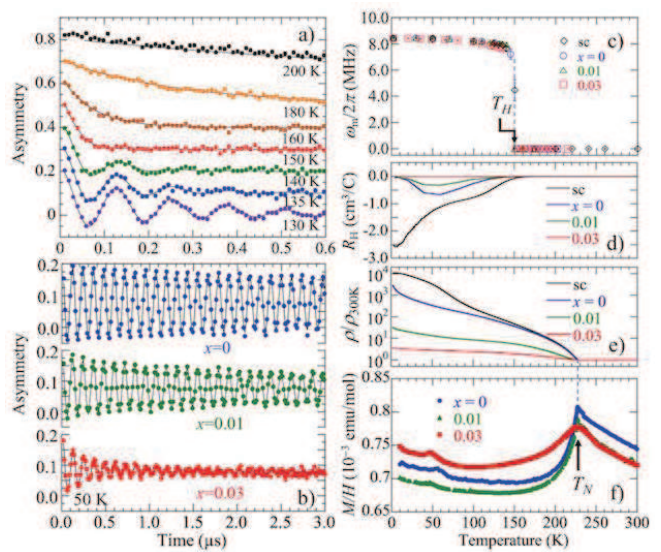}
			\caption{Typical ZF-$\mu$SR time spectra observed in (a) single-crystalline (sc-) \cdo\ at various temperatures (up to 0.5 $\mu$s) and (b) in \cdoro\ at 50 K for $x=0$, 0.01, and 0.03 (up to 3 $\mu$s). Temperature dependence of (c) spontaneous muon precession frequency deduced from the ZF-$\mu$SR spectra, (d) the Hall coefficient, (e) resistivity normalized at 300 K, and (f) magnetic susceptibility measured at 2 T, with $T_H$ and $T_N$ representing the characteristic temperatures.}
			\label{tspec}
	\end{center}
\end{figure}

As shown in Fig.\ref{tspec}a,  the \msr\ time spectra above $T_H$ exhibit depolarization described by a simple exponential damping without oscillation, 
\begin{equation}
 A(t)=A_0\exp(-\lambda_{\parallel}^* t),  \label{At0}
\end{equation}
where $A_0$ and $\lambda_\parallel^*$ are the initial asymmetry and the longitudinal depolarization rate, respectively.  Meanwhile, the spectra below $T_H$ are well represented by a single component of damping oscillation. More specifically, these spectra are described by the following simple form for the mosaic crystals (corresponding to the polycrystalline average),
\begin{equation}
      A(t)\simeq A_0[\frac{2}{3}\exp(-\lambda_\perp t)\cos\omega_{\rm m} t +\frac{1}{3}\exp(-\lambda_{\parallel} t)], \label{At1}
 \end{equation}
where $\omega_{\rm m}=\gamma_\mu B_{\rm m}$ is the spontaneous muon precession frequency, $\gamma_\mu$ is the muon gyromagnetic ratio ($= 2\pi \times 135.53$ MHz/T),  $B_{\rm m}$ is the internal local field probed by muon, and 
$ \lambda_\perp$, $ \lambda_\parallel$ are the transverse and longitudinal depolarization rate, respectively.  
$B_{\rm m}$ is predominantly determined by the vector sum of the magnetic dipolar fields from the local Os moments,  so that we have
\begin{equation}
B_{\rm m} =|\sum_j{\bf \hat{A}}_j{\bm \mu}_j|,\label{Bdip}
\end{equation}
 where ${\bf \hat{A}}_j$ is the dipole tensor and is expressed by 
\begin{equation}
{\bf \hat{A}}_j=A^{\alpha\beta}_j=(3\alpha_j\beta_j-\delta_{\alpha\beta}r_j^2)/r_j^5\quad(\alpha, \beta=x,y,z),\label{Adip}
\end{equation}
and the summation runs through the $j$-th Os$^{5+}$ moment ${\bm \mu}_j$ located at ${\bm r}_j=(x_j,y_j,z_j)$ from a muon site.  

It is noticeable in Figs.~\ref{tspec}b and \ref{tspec}c that muon depolarization is strongly influenced by mobile charge carriers. The Re substitution for Os leads to the increase of $\lambda_\perp$ with increasing $x$ below $T_H$ where no intrinsic carriers are presumed to exist (for $\lambda_\parallel$, see below).  Moreover, the disappearance of coherent muon spin precession above $T_H$ \cite{Koda:07} is in coincidence with the increase of the carrier density and/or the mobility as inferred from the Hall coefficient shown in Fig.~\ref{tspec}d. It is also clear in Figs.~\ref{tspec}e and \ref{tspec}f that $T_N$ remains unchanged upon Re substitution.

\begin{figure}[t]
	\begin{center}
\includegraphics[width=\linewidth]{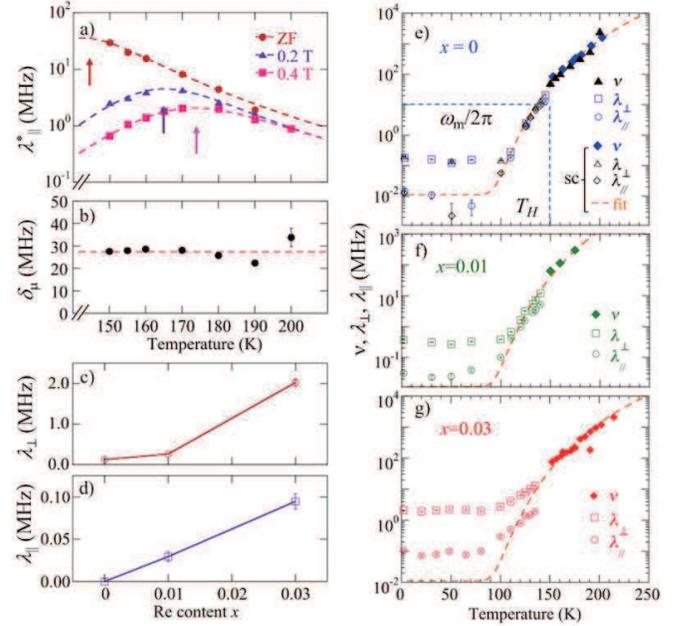}
			\caption{(a) Muon depolarization rate versus $T$ ($> T_H\simeq 150$ K) in sc-\cdo\ under a longitudinal field (LF) of 0, 0.2, and 0.4 T, and (b) muon-Os hyperfine parameter deduced from the data shown in (a) by fits using Eq.~(\ref{lmds}). Arrows indicate maxima of $\lambda_\parallel^*$ for the respective sets of data, and dashed curves are the least-square fits using Eqs.~(\ref{lmds}) and (\ref{Act}) in the text.  (c), (d) Muon depolarization rate ($\lambda_\perp$, $\lambda_\parallel$) vs.~Re content $x$ at 50 K ($<T_H$). (e)--(g) $\lambda_\parallel$, $\lambda_\perp$, and spin fluctuation rate ($\nu$) in \cdoro\ for the samples with respective Re content. Dashed curves show the result of least square fit for $\nu$ and $\lambda_\parallel$ in sc-\cdo.}
			\label{lmdnu}
	\end{center}
\end{figure}

The depolarization rate $\lambda_\parallel^*$ above $T_H$ deduced by curve-fit using Eq.(\ref{At0})  is shown in Fig.~\ref{lmdnu}a.  Here, a convincing evidence for the spin fluctuation as the origin of $\lambda_\parallel^*$ manifests in its dependence on the longitudinal magnetic field $B$, which is described by the following equation \cite{Miyazaki:16}, 
\begin{eqnarray}
\lambda_\parallel^*(B) &\simeq&\frac{2\delta_\mu^2\nu}{\nu^2+\omega_\mu^2}, \label{lmds}\\
\delta_\mu^2&=&\frac{1}{2}\gamma_\mu^2\mu_{\rm Os}^2\sum_{j,\alpha,\beta}(A^{\alpha\beta}_j)^2 
\end{eqnarray}
where $\delta_\mu$ is determined by the second moments of the dipole tensor ${\bf \hat{A}}_j$ [Eq.~(\ref{Adip}), with summation for $\alpha=x,y,z$, and $\beta=x,y$], $\mu_{\rm Os}=|{\bm \mu}_j|$, and $\omega_\mu=\gamma_\mu B$ is the muon precession frequency.  
It is noteworthy in Fig.~\ref{lmdnu}a that the peak of $\lambda_\parallel^*$ is clearly observed at a temperature $T_{\rm p}$ in the each set of data for a common $B$, indicating that a resonant enhancement of longitudinal spin relaxation occurs at $\nu(T_{\rm p}) = \omega_\mu$ as expected in Eq.(\ref{lmds}). The upward shift of $T_{\rm p}$ with $B$ indicates that $\nu(T)$ increases monotonically with temperature rise, thereby allowing us to deduce $\nu$ uniquely from Eq.(\ref{lmds}) that generally yields two solutions as a quadratic equation of $\nu$. 
The magnitude of $\delta_\mu$ was also determined experimentally by the $B$-dependence of $\lambda_\parallel^
*$ at each temperature (see Fig.~\ref{lmdnu}b), yielding  $\delta_\mu= 27.8(9)$ MHz as a mean value for $T_H\lesssim T<T_N$. 


The depolarization for $T< T_H$ observed under a zero field can be described as a part of the above-mentioned resonant relaxation, where $\delta_\mu$ and $\omega_\mu$ in Eq.~(\ref{lmds}) are predominantly determined by $B_{\rm m}$ to yield 
\begin{eqnarray}
 \lambda_\parallel&\simeq&\frac{2}{3}\left[\frac{2\omega_{\rm m}^2\nu}{\nu^2+\omega_{\rm m}^2}\right],\label{lmd}\\ \lambda_\perp&\simeq&\frac{\lambda_\parallel}{2}+\lambda_{\rm m},
\end{eqnarray}
where $\lambda_{\rm m}=\gamma_\mu\sqrt{\langle\mathit{\Delta}B_{\rm m}^2\rangle}$ with $\langle\mathit{\Delta}B_{\rm m}^2\rangle$ being the mean square of the quasi-static field distribution of $B_{\rm m}$. The prefactor 2/3 is derived from the fact that only 2/3 of the entire polarization is subject to decay by the fluctuation of $B_{\rm m}$ perpendicular to the initial muon polarization. When $\nu$ is much smaller than $\omega_{\rm m}$, Eqs.~(\ref{lmd}) are further approximated to yield $\lambda_\parallel\simeq4\nu/3$ and $\lambda_\perp\simeq\lambda_{\rm m}+2\nu/3$.  Here it must be stressed that $\lambda_\parallel$ is directly proportional to $\nu$, so that the non-zero $\lambda_\parallel$ for the samples with $x>0$ found in Fig.~\ref{lmdnu}d unambiguously indicates the presence of dynamical fluctuation induced by carrier doping.  Meanwhile, the non-linear $x$ dependence of $\lambda_\perp$ suggests contribution of Re-Re correlation to $\lambda_{\rm m}$ with increasing $x$ in determining the quasi-static randomness. 

The temperature dependence of $\nu$ ($T\gtrsim T_H$) is summarized in Figs.~\ref{lmdnu}e--\ref{lmdnu}f together with $\lambda_\parallel$ ($\simeq\nu$) and $\lambda_\perp$ for $T\lesssim T_H$.  It is noticeable that these parameters exhibit monotonic increase with temperature for $T\gtrsim100$ K. In particular, $\nu$ exceeds $10^9$ s$^{-1}$ for $T\gtrsim200$ K, which is unlikely to be induced by the self-diffusion of muon because the muon is suggested to be in a bonding state with ligand oxygen (i.e., forming an OH-bond, see Sec.~IV.A).  Thus the fluctuation can be reasonably attributed to the local Os spin fluctuation.

The loss of coherent muon spin precession above $T_H$ is attributed to the sharp increase of $\lambda_\parallel$ with increasing $\nu$ towards $\omega_{\rm m}= 5.2894\times10^7$ s$^{-1}$  (see Fig.~\ref{lmdnu}e) where Eq.(\ref{lmd}) exhibits maximum [known as ``$T_1$ minimum" in nuclear magnetic resonance (NMR)]. Thus $T_H$ is defined as the temperature where the condition $$\lambda_\perp\simeq\frac{\lambda_\parallel}{2}\simeq\frac{\omega_{\rm m}}{2\pi}$$ is satisfied. The relatively smooth change of $\nu$ in passing through $T_H$ is in line with the fact that the sublattice magnetization $|{\bm M}|$ ($\propto|{\bm B}_{\rm m}|$) exhibits only gradual change over the relevant temperature range.  Such a resonant depolarization process (which we call ``$T_2$ anomaly") does not occur for the magnon/spin-wave excitation because the frequency range ($\sim$10$^{12}$ s$^{-1}$) is far greater than $\omega_{\rm m}$ \cite{Nguyen:17}.

We note that the similar $T_2$ anomaly has been also suggested in the preliminary result of $^{17}$O-NMR measurements on single-crystalline \cdo, which is discussed in more detail in the Supplemental Material \cite{SM}.  

\section{IV. Discussion and Conclusion}
\subsection{A. Muon Site}
\begin{figure}[t]
	\begin{center}
\includegraphics[width=\linewidth]{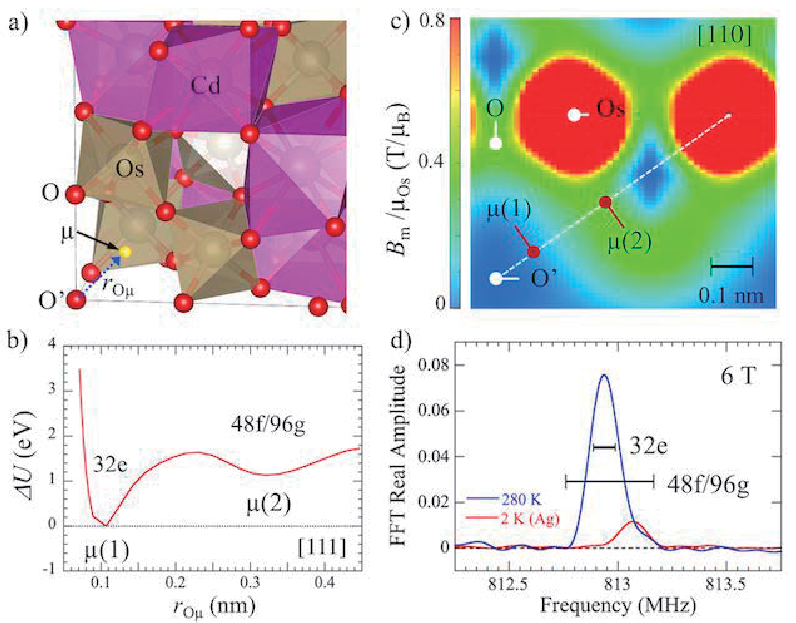}
\caption{(a) Crystal structure of \cdo\ with the arrow showing the possible muon position with distance $r_{\rm O\mu}$ from the O' site along $\langle111\rangle$ axis. (b) The total energy of \cdo\ hosting a hydrogen atom in the interstitial position evaluated by the DFT calculation,  where the position $\mu(1)$ corresponding to the 32e site (Wycoff position) exhibits a minimum. A local minimum $\mu(2)$ is located near those with 48f/96g sites. (c) A contour map of the local magnetic field calculated for the $\langle110\rangle$ plane, where the Os moment size of 1$\mu_B$ is assumed. The labels $\mu(i)$ indicate the local potential minima shown in (b). (d) The fast Fourier transform of the \msr\ spectra under a transverse field of 6 T, where the bars indicate linewidth expected for the $\mu(i)$ positions. The signal from silver sample holder (Ag) is also observed at +94 ppm (2 K).  }
			\label{siteL}
	\end{center}
\end{figure}

Since muon behaves as a pseudo-hydrogen in matter, the variation in the total electron energy upon hydrogen inclusion estimated by the density functional theory (DFT) calculation serves as a useful guide in narrowing down the candidate muon sites.  The most probable site is then inferred from the three experimentally observable quantities, i.e., the muon Knight shift and associated linewidth in the paramagnetic phase ($T>T_N$), the hyperfine parameter $\delta_\mu$ deduced from the field dependence of $\lambda_\parallel^*$ for $T_H<T<T_N$, and the internal magnetic field $B_{\rm m}$ in the AF ground state ($T\ll T_H$).  Following the suggestion from the DFT calculation for the total electron energy using the VASP code package \cite{VASP}, we examined positions around the two potential minima, $\mu$(1) (near the Wycoff position 32e of space group $Fd\bar{3}m$) and $\mu$(2) (near 48f/96g) positions along $\langle111\rangle$ direction shown in Figs.~\ref{siteL}a,b [crystal coordination (0.062,0.062,0.062) and (0.185,0.185,0.185), corresponding to the distance from the oxygen (O') site $r_{\rm O\mu}=0.11$ and 0.32 nm, respectively]. 

It is empirically well established that muons in transition metal oxides exhibit a tendency of forming local bound states with ligand oxygen at their interstitial sites \cite{Cox:06}. This has a semi-quantitative basis that hydrogen impurities exhibit a similar tendency in non-metallic compounds \cite{Klc:02,Walle:03}.  The suggested global minimum of the total energy at the $\mu$(1) position is indicative of such a tendency, as the O'H distance is close to the typical OH bond length ($\sim$0.1 nm).

${\bm T}>{\bm T_N}$:---It is suggested from the \msr\ the linewidth observed in the muon Knight shift measurements (under a transverse field of 6 T applied parallel with $\langle111\rangle$, see Fig.~\ref{siteL}d) that the muon position at 280 K ($>T_N$, paramagnetic semimetal phase) is located near the 32e site, because these predicted for the sites within the Os tetrahedra (48f or 96g) far exceed the experimental observation.  

\begin{figure}[t]
	\begin{center}
\includegraphics[width=\linewidth]{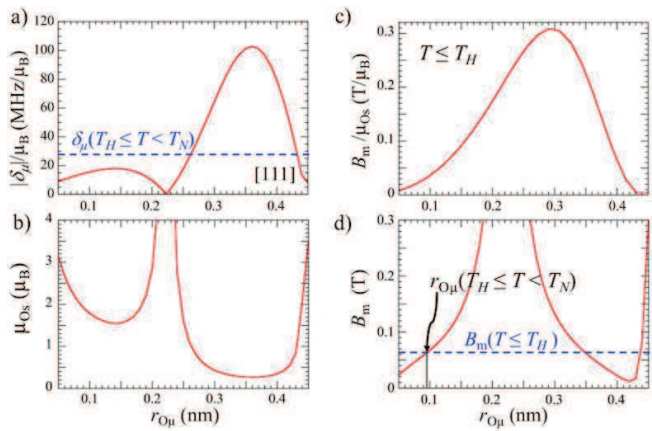}
\caption{(a) Muon-electron hyperfine parameter $|\delta_\mu|$ per Os moment $\mu_{\rm Os}$ in the paramagnetic phase calculated for the position $r_{\rm O\mu}$ along the $\langle111\rangle$ direction. (b) The effective size of $\mu_{\rm Os}$ obtained by dividing experimental $\delta_\mu$ by $|\delta_\mu|/\mu_{\rm Os}$ in (a). (c) The local magnetic field $B_{\rm m}$ per $\mu_{\rm Os}$ calculated along the $\langle111\rangle$ axis quoted from Fig.~\ref{siteL}c. (d) The $r_{\rm O\mu}$ dependence of the calculated $B_{\rm m}$ obtained as the product of (b) and (c), where $B_{\rm m}$ coincides with the experimental value (62.1 mT) for $T\le T_H$ at $r_{\rm O\mu}\simeq0.094$ nm, which is in good agreement with $r_{\rm O\mu}$ for $T<T_H$.}
			\label{siteH}
	\end{center}
\end{figure}

${\bm T_H}\lesssim{\bm T}<{\bm T_N}$:---In this temperature range, the muon position is examined by the consistency of $B_{\rm m}$ (as a mean value) estimated by $\delta_\mu$ with that observed below $T_H$ for various $r_{\rm O\mu}$, because the present \msr\ result suggests that $B_{\rm m}$ remains well established at least around $T_H$.  The calculated $|\delta_\mu|/\mu_B$ versus  $r_{\rm O\mu}$ is shown in Fig.~\ref{siteH}a, from which we can evaluate the possible Os moment size for the corresponding $\delta_\mu({\rm exp.})$, 
\begin{equation}
\mu_{\rm Os}=\frac{\delta_\mu({\rm exp.})}{|\delta_\mu|}\mu_B
\end{equation}
which is shown in Fig.~\ref{siteH}b.  Using the calculated $B_{\rm m}/\mu_{\rm Os}$ along the $\langle111\rangle$ axis in Fig.~\ref{siteH}c (which is extracted from Fig.~\ref{siteL}c), one can estimate $B_{\rm m}=(B_{\rm m}/\mu_{\rm Os})\times\mu_{\rm Os}$ versus $r_{\rm O\mu}$ as shown in Fig.~\ref{siteH}d. $B_{\rm m}$ coincides with experimental value at $r_{\rm O\mu}\simeq0.094$ nm in reasonable agreement with that estimated for $T\lesssim T_H$. (Another candidate position with $r_{\rm O\mu}\simeq0.34$ is close to the 48f/96g sites and thereby it can be excluded.)

${\bm T}\lesssim{\bm T_H}$:---The muon site for the AI phase is inferred from the magnitude of $B_{\rm m}$ associated with the AIAO magnetic order, where $B_{\rm m}$ is predominantly determined by the magnetic dipolar field from the static Os moments given by Eq.~(\ref{Bdip}).
 Fig.~\ref{siteL}c shows a contour map of $B_{\rm m}$ for the $\langle110\rangle$ plane including the O' site  calculated for $\mu_{\rm Os}=1\mu_B$. 
The curve-fit of the \msr\ spectra for $T\lesssim T_H$ indicates that $\omega_{\rm m}/2\pi$ converges to 8.4184(4) MHz with decreasing temperature, yielding  $B_{\rm m}=62.110(3)$ mT. This corresponds to the muon position at  $r_{\rm O\mu}=0.127$ nm from the O' site along the $\langle111\rangle$ direction [crystal coordination (0.072,0.072,0.072)], which is in reasonable agreement with the $\mu$(1) site suggested to be the most stable position for the interstitial hydrogen showing the global minimum of the total energy. 

Thus, it is concluded that muon occupies nearly identical sites throughout all the electronic phases in \cdo\ despite the change in the electronic property, indicating that the O$\mu$ bonded state is quite stable in this compound.
This is corroborated by the relatively large energy gain associated with the $\mu(1)$ position shown in Fig.~\ref{siteL}b, where the potential barrier for migration to the next nearest site exceeds $\sim$ 1.6 eV. Even the relatively large zero-point energy for muon ($\simeq0.1$--0.2 eV) would not change the situation, making it quite unlikely that muon diffusion occurs with the observed activation energy ($E_a\simeq0.21$ eV) within the relevant temperature range.

\subsection{B. Magnetoelastic polaron model}
\begin{figure}[t]
\begin{center}
\includegraphics[width=0.95\linewidth]{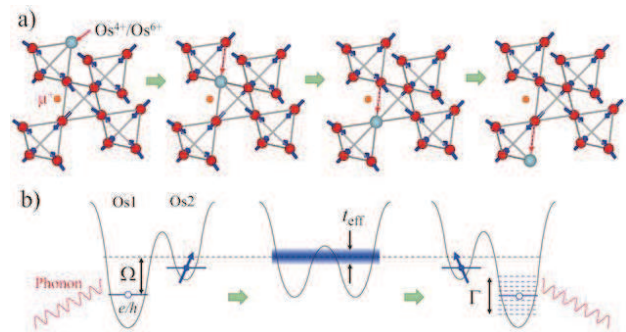}
	\caption{Schematic illustrations of magnetoelastic polaron motion showing a) sequential spin configurations and b) their energetics. In a), the Os$^{4+}$ (electron) or Os$^{6+}$ (hole) state is presumed to be in the low spin state ($S=0$) of the $t_{2g}$ multiplet (opaque ball), which locally modulates AIAO spin configuration successively in moving from top to the bottom on an Os chain (left to right). In b), phonons are involved to adjust the magnetoelastic polaron energy ($\Omega$) for tunneling from the Os1 to Os2 site (with an effective matrix $t_{\rm eff}$). These phonons also cause potential fluctuation ($\Gamma$) to reduce the tunneling probability.}
	\label{sol}
\end{center}
\end{figure}

Considering that the density of states near $E_F$ comes from the Os 5$d$ orbitals, the hopping of carriers is expected to induce temporal modulation of the local spin configuration due to the fluctuation of Os valence, thus leading to the fluctuation of $B_{\rm m}$ observed by muons at the nearest neighboring sites. More specifically, the magnetic transitions may occur between $5d^3$ ($S=3/2$) and $5d^2$ or $5d^4$ (the $S=0$ low spin state expected for the $5d$ metals), which leads to a local three-in or three-out spin configuration as schematically illustrated in Figs.~\ref{sol}a. The doping of itinerant holes by Re substitution would lead to a similar consequence, dominating the fluctuation below $T_H$. The fluctuation rate observed by muon is then given by $$\nu\simeq\sigma_{\rm c} v n,$$ where $\sigma_{\rm c}$ is the cross section for carriers to interact with the Os ion neighboring to the muon, $v$ is the carrier velocity, and $n$ is the carrier density (we disregard the difference between electrons and holes \cite{Hiroi:15}). Given that the Fermi level is situated at the center of the energy gap (see Fig.~\ref{phasdia}), the temperature dependence of $n$ in the AI phase is given by the equation $n(T)\simeq n_0\exp[-\Delta_D(T)/k_BT]$, where $n_0$ is the carrier density in the semimetallic phase which is theoretically estimated to be $1.27\times10^{21}$ cm$^{-3}$ \cite{Harima}. 
However, it turns out that the curve-fit analysis by the relation $\nu(T)\propto\exp(-E_{\rm a}/k_BT)$ yields $E_{\rm a}=0.210(2)$ eV which is significantly greater than $\Delta_D(T)$ ($\simeq0.050$--0.075 eV \cite{Sohn:15}), indicating the need for considering additional interaction that reduces $v$ and/or $n$.

Here, the presence of strong spin-phonon coupling \cite{Sohn:17} naturally leads to the possibility that
the large $E_{\rm a}$ is associated with the formation of small polarons in the AI phase where charge screening by free carriers is diminished.  The relevant quasiparticle may be called magnetoelastic polaron, as the carrier motion accompanying spin-flip requires extra cost of energy determined by the magnetoelastic interaction.   As shown in Fig.~\ref{sol}b, the small polarons exhibit hopping motion described by $v\simeq a t_{\rm eff}^2/\hbar\Gamma(T)$, where $a$ is the distance of carrier hopping,  $t_{\rm eff}$ is the effective tunnel matrix, and $\Gamma(T)$  is the final state potential fluctuation by phonons \cite{Holstein:59}.  The polaron is called ``small" when $t_{\rm eff} =t_0 \exp[-S(T)]\ll \Gamma(T)$, where $t_0$ and $S(T)$ refer to the transfer matrix of carriers in the rigid lattice and the renormalization factor due to the lattice relaxation, respectively.  $S(T)$ converges to $S(0)=5\Omega/2k_B\Theta_D$ for $T\lesssim2\Theta_D/5 =160$ K [with $\Theta_D\simeq400$ K being the Debye temperature \cite{Mandrus:01})] , whereas it is governed by thermal activation for $T>T_H$ so that $S(T)\simeq\Omega/k_BT$ with $\Omega$ predominantly determined by the energy required to restore the translational symmetry via multi-phonon process. The good agreement between $T_H$ and $2\Theta_D/5$ provides further evidence that $T_H$ is the characteristic temperature for the small polaron hopping.

Considering that $\Gamma(T)=\Gamma_0\exp[-S(T)]$ \cite{Holstein:59}, 
the entire temperature dependence of $\nu$ may be then described by the equation
\begin{equation}
\nu(T) \simeq\sigma_{\rm c} v_0 n_0 \exp\left[-\frac{\Delta_D+\Omega}{k_BT}\right]+\nu_x,\label{Act}
\end{equation}
where $v_0=at^2_0/\hbar\Gamma_0$, and $\nu_x$ is for the fluctuation induced by carrier doping. A curve-fit using Eq.~(\ref{Act}) for the data of sc-\cdo\ sample yields $\sigma_{\rm c} v_0 n_0=4.79(46)\times10^{13}$ s$^{-1}$ and $\Omega=E_{\rm a}-\Delta_D(T)\simeq0.14$--0.16 eV. The corresponding $\nu(T)$ curve is shown in Figs.~\ref{lmdnu}e--\ref{lmdnu}f for comparison with the data in polycrystalline samples. 
Using a coarse estimation that $\sigma_{\rm c}\simeq2\times10^{-15}$ cm$^{2}$ (the cross section of an Os tetrahedron), we have $v_0\simeq10^{7}$ cm/s which is in reasonable agreement with the Fermi velocity obtained by the LDA calculation ($\simeq5.9\times10^6$ cm/s \cite{Singh:02}). The magnitude of $\Omega$ is consistent with the energy required for the multiphonon process involving the spin-coupled phonons \cite{Sohn:17}.  The energy cost for the adiabatic spin-flip ($\Delta_M$, which comprises a part of $\Omega$) would be similar to the magnetic transitions observed by the resonant inelastic X-ray scattering \cite{Calder:16,Bogdanov:13}, where only the lower energy part ($10^1$--$10^2$ meV, $<E_{\rm a}$) is relevant. 

The absence of anomaly around $T_H$ and/or $T^*$ in the magnetic susceptibility (see Fig.~\ref{tspec}f) and in the RXMS amplitude \cite{Yamaura:12}  is explained by the well localized (which is soliton-like, because of the Ising anisotropy) and incoherent characters of the magnetic fluctuation that result in the reduction of the order parameter only by a factor determined by the carrier density, $\mathit{\Delta}M/M\simeq n(T)/n_0$. Meanwhile, the discrepancy between $T_N$ and $T^*$ implies that the AIAO magnetic order is driven by the local (${\bm q}=0$) energy gain of Os magnetic moments irrespective of the Fermi surface topology. 

The evidence for the small-polaron conduction in the AI phase implies the importance of magnetoelastic coupling in understanding the transport properties in the semimetallic phase as well. Provided that the MI transition progresses concomitantly with $U$ becoming dominant over $T$ at low temperatures, $n(T)$ would exhibit only modest reduction around $T_N$. Considering the distinct correlation between $T_N$ and the increase of $\rho(T)$, it is tempting to speculate that the magnetoelastic coupling serves to reduce the mobility by forming a polaronic state [e.g., a large polaron that satisfies the relation $t_{\rm eff}\gg\Gamma(T)$] in the AS phase. This is corroborated by the remarkable concordance in the $T$ dependence between the plasma frequency \cite{Sohn:15} and the frequency of the spin-coupled phonons \cite{Sohn:17}.  


In summary, we found a novel spin-charge-phonon fluctuation in the AIAO pyrochlore magnet \cdo, where the spin fluctuation is driven by the conduction of carriers throughout semimetal (AS) and insulator (AI) phases. The characteristic energy scales involved in the fluctuation suggests the conduction of magnetoelastic polarons in the AI phase.  This in turn suggests need to reconsider the electronic properties of the AS phase from the viewpoint of the MI transition at finite temperatures.

\section*{Acknowledgements} We would like express our thanks to the staff of TRIUMF and PSI for their technical support during the \msr\ experiment, and to H. Lee for his help on the DFT calculation.  We also appreciate stimulating discussion with H. Shinaoka, M. Udagawa, Y. Motome, M. Ogata, and Y. Kuramoto.  This work was partially supported by Condensed Matter Research Center, IMSS, KEK.

\end{document}